\documentclass[twoside]{article}
\usepackage[a4paper]{geometry}
\usepackage[utf8]{inputenc} % ou \usepackage[utf8]{inputenc}
\usepackage[T1]{fontenc} % ou \usepackage[OT1]{fontenc}
\usepackage{RR}
\usepackage{hyperref}
\usepackage[english,american,french]{babel}
\def\og{\leavevmode«~}
\def\fg{\ifdim\lastskip>\z@\unskip\fi~»}
\RRNo{8100}
\RRdate{October 2012}
\RRauthor{% les auteurs
François Pellegrini\thanks{LaBRI \& INRIA Bordeaux Sud-Ouest. \texttt{francois.pellegrini@inria.fr}}%
  \and
Sébastien Canevet\thanks{Université de Poitiers, IUT GEA, 8~rue Archimède, 79000 Niort, France. \texttt{sebastien@canevet.net}}%
}
\authorhead{Pellegrini \& Canevet}
\RRtitle{Le droit du numérique~: une histoire à préserver}
\RRetitle{Le droit du numérique~: une histoire à préserver}
\titlehead{Digital Law: a History to Preserve}
\RRresume{Bien que l'histoire de l'informatique soit récente, cette
  discipline pose des problèmes de conservation inédits. Ceux-ci sont
  amplifiés par des questions d'ordre juridique, le droit du numérique
  constituant en lui-même un objet d'étude dont l'histoire a également
  sa place dans un musée de l'informatique. L'objectif de cet article
  est de proposer un tour d'horizon de l'évolution du droit du
  numérique, tant du point de vue historique que de son impact sur la
  conservation et la présentation des pièces.}
\RRabstract{Although the history of informatics is recent, this field
  poses unusual problems with respect to its preservation. These
  problems are amplified by legal issues, digital law being in itself
  a subject matter whose history is also worth presenting in a
  computer science museum. The purpose of this paper is to present a
  quick overview of the evolution of law regarding digital matters,
  from an historical perspective as well as with respect to the
  preservation and presentation of the works.}
\RRmotcle{droit, numérique, logiciel, droit d'auteur, licence, format,
  interopérabilité, données personnelles, Internet}
\RRkeyword{law, digital, software, copyright, license, format,
  interoperability, privacy, Internet}
\RRprojet{Bacchus}
\RCBordeaux % centre de recherche Bordeaux - Sud Ouest
\begin{document}
\makeRR   % cas d'un rapport de recherche
%% \makeRT % cas d'un rapport technique.
%% a partir d'ici, chacun fait comme il le souhaite

\section{Introduction}

Le logiciel a un rôle à part dans l'histoire humaine~: c'est le premier
outil mécanisé qui soit une extension de notre esprit plutôt que de
notre corps. Alors que la machine, moteur et objet de la révolution
industrielle, permet à une personne de mettre en jeu une puissance
physique plus grande que celle de son corps, le logiciel, moteur et
objet de la révolution numérique, permet à cette même personne de
traiter l'information avec une puissance supérieure à celle de son
esprit.

Confronté à l'irruption des technologies numériques, le législateur
a été amené à questionner les fondements du droit autant qu'à organiser
les usages nouveaux que permettaient la technique. Identité numérique,
statut des données personnelles (avec la naissance de la CNIL, suscitée
par la capacité de traitement informatisé), statut et neutralité d'Internet,
interopérabilité, protection juridique du logiciel, variété des licences
logicielles et des modèles économiques, mutation du droit d'auteur
et du droit de la presse, etc., tels sont les nombreux chantiers,
pour la plupart encore ouverts, auxquels l'internationalisation des
échanges numériques ajoute une complexité supplémentaire.

Nous pensons que la sauvegarde et la présentation au public du patrimoine
lié au numérique doit inclure ces aspects juridiques, afin d'illustrer
l'étendue des bouleversements apportés à la société par l'irruption
du numérique, et la manière dont ils ont été appréhendés.

Dans cet article, nous nous attacherons à identifier les principaux
domaines du droit qui ont été impactés par la révolution numérique.
Nous évoquerons pour chacun d'entre eux les textes et dates clés qui
permettent de les replacer dans une perspective historique. Ce premier
travail n'a pas pour ambition d'être exhaustif, mais plutôt d'aider
les spécialistes de la muséographie à créer de nouveaux liens entre
les différents témoins de l'histoire numérique, en dehors du seul
angle de l'évolution de la technique. Nous nous focaliserons plus
spécifiquement sur quelques secteurs, dont celui d'Internet ainsi
que celui du logiciel.

\section{Quel droit pour le logiciel~?}

Tant que les ordinateurs n'étaient construits qu'en quelques exemplaires,
l'activité de développement de programmes restait fortement liée à
l'activité de conception des machines. Son coût était dilué dans ceux
de conception et de maintenance des matériels. Ce n'est qu'à partir
de l'existence d'une base installée suffisante d'ordinateurs de même
type%
\footnote{L'UNIVAC~I de la société Remington Rand, commercialisé à partir de
juin~1951, fut le premier ordinateur produit de façon industrielle
et fut vendu à 46 exemplaires.%
} que l'activité de programmation a pu être vue comme une activité
économique indépendante, découplée de la fabrication de l'ordinateur
lui-même.

La prise de conscience de la nature spécifique du logiciel, puis de
son existence en tant que produit autonome, a trouvé son aboutissement
à la fin des années~1960. À cette époque, la société IBM accaparait
quatre-vingt-dix pour cent du marché mondial de l'informatique. Elle
distribuait gratuitement son système d'exploitation, ses logiciels
d'usage général et leurs codes sources, comme autant de fournitures
annexes à ses ordinateurs, au même titre que les manuels d'utilisation.

Face à l'arrivée de concurrents souhaitant commercialiser des matériels
compatibles avec ses propres ordinateurs, et parce que la divulgation
du code source de son système d'exploitation facilitait le travail
de ses concurrents, IBM décida de ne plus le leur fournir. Les dits
concurrents contactèrent alors le Département de la justice afin d'entamer
des poursuites contre IBM pour abus de position dominante, sur le
fondement du \foreignlanguage{american}{\textit{Sherman Act}}.

Afin de contrer cette action, lancée en janvier~1969, IBM annonça
dès juin~1969 une politique d'\og \foreignlanguage{american}{\textit{unbundling}} \fg{}
(\og dégroupage \fg{}), c'est-à-dire de fourniture et de facturation
séparée des matériels et des logiciels%
\footnote{IBM avait évalué forfaitairement à trois pour cent le prix des logiciels,
de la formation et de l\textquoteright{}assistance technique, les
quatre-vingt-dix-sept pour cent restants revenant encore, noblesse
oblige, au matériel.%
}. Le droit des affaires n'étant souvent que la représentation d'une
réalité comptable par d'autres moyens, cette décision eut pour conséquence
l'apparition des premières Conditions générales de vente.

L'affaire n'en resta pas là, et une cascade de procès secondaires
amena à faire condamner~IBM devant la Cour suprême en~1972. Cette
décision obligea IBM d'une part a mettre en œuvre de façon effective
la facturation séparée son matériel et de ses logiciels, et d'autre
part à devoir fournir ces derniers sans discrimination, y compris
à ses concurrents sur le segment des matériels%
\footnote{Quant aux investigations initiales du Département de la justice, elles
suivirent leur cours sous la forme d'une guerre de tranchées juridique,
qui ne se conclut qu'en~1982 (!) par un classement de l'affaire,
le procès étant jugé sans objet (\og \foreignlanguage{american}{\textit{without
merit}} \fg{}) au vu de l'évolution de la situation.%
}.

Ces événements marquent ainsi la date de naissance de l'industrie
du logiciel. Ils conduisirent naturellement à s'interroger sur la
protection de ce dernier par le droit, ainsi que sur le statut de
ses créateurs.

\subsection{Le choix du droit d'auteur}

Les logiciels sont tout à la fois œuvres de l'esprit et biens substituables%
\footnote{On appelle substituable un bien qui peut remplacer ou être remplacé
par un autre bien pour répondre à un même besoin. La substituabilité
d'un bien ou d'un service peut dépendre de facteurs externes. Par
exemple, certains produits peuvent devenir substituables au pétrole
en raison de son renchérissement~; le train et l'avion peuvent être
des services substituables si la géographie le permet, etc.%
} à usage industriel. Pour assurer leur protection, trois possibilités
s'offraient au législateur français~: élaborer un droit autonome
spécifique au logiciel, rattacher la protection du logiciel au droit
d'auteur ou la rattacher à celle des brevets.

La voie du droit spécifique était sans doute la solution la plus adaptée
du point de vue théorique. En revanche, elle présentait le notable
inconvénient de n'être qu'une protection franco-française, sans reconnaissance
ni efficacité à l'international, ce qui eut pris des années à mettre
en place.

La voie de la protection par le brevet ayant été écartée assez vite,
parce qu'inadaptée, c'est en définitive la voie du rattachement au
droit d'auteur qui a été privilégiée. Elle présentait le notable avantage
de permettre au logiciel de bénéficier immédiatement de l'ensemble
des conventions et traités internationaux protégeant le droit d'auteur,
notamment la convention de Berne de~1886.

La jurisprudence devança même le législateur, puisque les premières
décisions précèdent l'intégration du droit du logiciel dans le Code
de la propriété intellectuelle (CPI). La jurisprudence fondatrice,
la célèbre affaire Babolat c/~Pachot, a donné lieu à une première
décision dès~1982%
\footnote{Si l'affaire \og Pachot \fg{} ne vit son terme que le 7~mars 1986,
par une décision de la Cour de cassation (Cass. Ass. Plén., 7~mars
1986, n\textdegree{}~83-10477, Babolat c/~Pachot), celle-ci fut
précédée de plusieurs jugements et arrêts antérieurs~: jugement du
18~novembre 1980 du Tribunal de commerce (15\textsuperscript{è}
chambre), et arrêt du 2~novembre 1982 de la Cour d\textquoteright{}appel
de Paris (4\textsuperscript{è} chambre).

C'est à la Cour de cassation que revient le mérite d'avoir formalisé
la notion d'\og apport intellectuel \fg{} pour qualifier ce qui
relève de l'\og originalité \fg{} de l'apport de l'auteur dans le
domaine artistique.%
}, alors que ce n'est que par le biais de la loi du 3~juillet 1985%
\footnote{Loi n\textdegree{}~85-660 du 3~juillet 1985 relative aux droits
d'auteur et aux droits des artistes-interprètes, des producteurs de
phonogrammes et de vidéogrammes et des entreprises de communication
audiovisuelle.%
} que le Parlement accueillit explicitement le logiciel au sein du
droit d'auteur. Pour tenir compte des spécificités de l'œuvre logicielle,
plus industrielle qu'artistique, la loi de~1985 amenda cependant
le droit d'auteur applicable aux programmes informatiques, d'une façon
que nous préciserons plus bas.

\medskip{}

Aux États-Unis, la \foreignlanguage{american}{\textit{Commission on
New Technological Uses of Copyrighted Works}} (« Commission des nouveaux
usages technologiques des travaux protégés par le copyright ») fut
créée en~1974 pour contribuer à la modernisation de la loi étatsunienne
sur le copyright suite à l'émergence des nouvelles technologies de
l'époque~: le photocopieur et l'ordinateur. Elle recommanda que le
logiciel puisse être protégé par le copyright. Cette idée fut mise
en œuvre en~1980 par le Congrès étatsunien, qui étendit la protection
du Copyright Act aux logiciels, en insérant au paragraphe~111 de
ce texte une définition des \og \foreignlanguage{american}{\textit{computer
programs}}\fg{} les protégeant au même titre qu'une œuvre littéraire.
La jurisprudence emboîta rapidement le pas au législateur, notamment
par le célèbre arrêt Apple v.~Franklin%
\footnote{Apple Computer, Inc. v. Franklin Computer Corporation. U.S. Court
of Appeals Third Circuit, 30 août 1983 - 714 F.2d 1240, 219 USPQ 113.%
}.

\medskip{}

Tous les pays confrontés à la question de la protection des programmes
d'ordinateur, incités par ces premiers mouvements, conclurent eux
aussi en faveur de l'intégration du logiciel dans le droit d'auteur.
Ce processus s'acheva en Europe avec la Directive du Conseil du 14~mai
1991 sur la protection des programmes d'ordinateurs%
\footnote{Directive 91/250/CEE du Conseil des communautés européennes, du 14
mai 1991, concernant la protection juridique des programmes d'ordinateur.
Journal officiel n\textdegree{} L~122 du 17/05/1991, p.~0042~-
0046.%
}, qui harmonisa cette protection au sein de l'Union européenne. Cette
directive fut transposée en France par la loi du 10~mai 1994%
\footnote{Loi no. 94-361 du 10~mai 1994 portant mise en œuvre de la directive
(C.E.E.) n\textdegree{}~91-250 du Conseil des communautés européennes
en date du 14~mai 1991 concernant la protection juridique des programmes
d'ordinateur et modifiant le CPI.%
}, qui modifia certaines des dispositions de la loi de~1985.

\subsection{Adaptation du droit d'auteur au logiciel}

Le logiciel est donc protégé par le droit d'auteur, au même titre
qu'un livre ou un tableau. Cependant, son caractère et sa finalité,
plus utilitaires qu'artistiques, ont conduit le législateur a adapter
le droit d'auteur à ses spécificités. Il s'est agi également de tenter
d'assurer un équilibre entre les droits des auteurs, ceux de leurs
employeurs et ceux des utilisateurs. Tant les droits patrimoniaux
que les droits extra-patrimoniaux ont donc été amputés par rapport
à ceux qui s'appliquent aux autres types d'œuvres. Les principales
dispositions le concernant ont été historiquement introduites par
le titre~V de la loi de~1985, intitulé~: \og Des logiciels \fg{}.

Tout d'abord, prenant acte du caractère industriel de la production
logicielle, les droits patrimoniaux concernant un logiciel réalisé
par un salarié ou fonctionnaire dans l'exercice de ses fonctions sont
systématiquement dévolus à l'employeur. Cette disposition rompit avec
le droit d'auteur traditionnel, centré sur l'auteur, qui disposait
à l'époque que les droits patrimoniaux de l'auteur de l'œuvre non
logicielle devaient faire l'objet d'un acte de cession explicite,
la simple existence du contrat de travail ne suffisant pas à transférer
à l'employeur les droits patrimoniaux sur l'œuvre créée. Sur ce point,
la loi DADVSI%
\footnote{Loi n\textdegree{}2006-961 du 1\textsuperscript{er}~août 2006 relative
au droit d'auteur et aux droits voisins dans la société de l'information.%
} de~2006 a finalement aligné le statut de toute œuvre créée par le
salarié ou le fonctionnaire dans l'exercice de sa mission sur celui
de l'œuvre logicielle, reflétant l'industrialisation de l'ensemble
de la production intellectuelle.

Également en rupture avec le statut traditionnel de l'auteur, la Loi
de~1985 a autorisé que la cession des droits sur une œuvre logicielle
puisse se faire pour un montant forfaitaire, et non sur la base de
royalties dépendant du nombre d'exemplaires vendus. Il s'agissait
de prendre acte de la réalisation de prestations à façon, le logiciel
réalisé en sous-traitance pouvant être utilisé ou revendu en un nombre
quelconque d'exemplaires par son commanditaire.

De même, afin que chaque licence puisse être facturée, l'exception
de copie privée fut supprimée pour les logiciels, et remplacée par
le droit d'effectuer une copie de sauvegarde, celle-ci ne pouvant
être utilisée qu'en cas de défaillance du support original.

Une disposition originale de la loi de~1985 fut de réduire la durée
des droits patrimoniaux sur les œuvres logicielles à 25 ans, au lieu
de 50~ans après la date de décès du dernier auteur survivant. Le
législateur prenait ainsi acte de l'obsolescence rapide des produits
dans ce secteur, du fait de leur substituabilité, et ouvrait la porte
à l'existence d'un domaine public logiciel vivant. Cette exception
fut malheureusement supprimée lors de la transposition de la directive
91/250/CE%
\footnote{Il peut cependant exister une fenêtre d'opportunité pour les logiciels
antérieurs à~1969, s'ils ont été publiquement diffusés en tant qu'œuvres
de domaine public avant la promulgation des décrets d'application
de la loi de~1994.%
}. La durée des droits patrimoniaux des logiciels, désormais soumise
au régime général, fut portée à 70~ans après le décès de l'auteur
par la loi de~1997%
\footnote{Loi n\textdegree{}~97-283 du 27~mars 1997 portant transposition
dans le Code de la propriété intellectuelle des directives du Conseil
des Communautés européennes n\textdegree{}~93/83 du 27~septembre
1993 et n\textdegree{}~93/98 du 29~octobre 1993. Dans le cas des
œuvres collectives, il s'agit d'une durée de 70~ans comptée après
le 1\textsuperscript{er} janvier suivant la date de parution initiale.%
}.

\subsection{La diversité des licences}

Avant même le rattachement effectif du logiciel au droit d'auteur,
les éditeurs de logiciels ont affiné un corpus de clauses contractuelles
afin d'empêcher l'usage de ceux-ci sans contrepartie. Ces dispositions,
initialement issues du droit des contrats industriels (fourniture
de produits et de prestations), furent reformulées selon les dispositions
du droit d'auteur une fois la protection du logiciel par ce dernier
acquise.

Aux États-Unis, le rattachement du logiciel au droit d'auteur permit
d'y remplacer le contrat par la \og \foreignlanguage{american}{\textit{license}} \fg{},
cette dernière y représentant un outil juridique bien plus intéressant
que le premier. Du fait que les États-Unis sont un pays fédéral, le
droit des contrats peut s'y appliquer différemment d'un État à l'autre,
alors que la \foreignlanguage{american}{\textit{license}}, qui relève
du droit fédéral, est d'une interprétation uniforme.

En France, où cette distinction n'est pas de mise, les \foreignlanguage{american}{\textit{licenses}}
sont considérées par le juge sous l'angle contractuel. On les y qualifie
de \og licences \fg{}, bien que ce terme désigne en droit national
un objet juridique différent, plus apparenté aux conventions collectives.
Ceci n'a pas empêché les tribunaux nationaux de les recevoir favorablement.

\medskip{}

Le modèle économique dominant des éditeurs de logiciels consiste à
vendre en de multiples exemplaires un même logiciel, dont le coût
de fabrication est (au moins partiellement) mutualisé entre les acheteurs.
Le prix de vente unitaire est alors inférieur au coût de développement
à façon qu'aurait eu à supporter individuellement chaque client.

Les \foreignlanguage{american}{\textit{licenses}} qui sous-tendent
ce modèle, que nous appellerons \og privatives%
\footnote{Nous préférons ce terme au terme de \og propriétaire \fg{} couramment
usité. Nous l'entendons presque dans le sens ou l'entend le droit
immobilier lorsqu'il distingue les \og parties privatives \fg{},
réservées au seul usage du copropriétaire, des parties communes qui
peuvent être utilisées par l'ensemble des copropriétaires et des usagers
du bien immobilier. Le terme \og propriétaire \fg{} nous semble
inapproprié car il peut faire accroire que seules ces \foreignlanguage{american}{\textit{licenses}}
relèvent de la \og propriété intellectuelle \fg{}, à l'opposé des
\foreignlanguage{american}{\textit{licenses}} \og libres \fg{} que
nous présenterons plus bas (et que certains appellent à tort \og libres
de droits \fg{}). Ceci est inexact, puisque toutes les \foreignlanguage{american}{\textit{licenses}}
s'appuient sur le droit d'auteur pour concéder des droits et~/ ou
des devoirs à leurs récipiendaires.%
} \fg{}, ont en commun de réserver l'intégralité des droits au titulaire
de ceux-ci. Même le droit d'usage peut en effet faire l'objet de restrictions%
\footnote{La créativité des éditeurs est en la matière sans limite. Par exemple,
Microsoft avait inséré dans les conditions générales d'utilisation
de son logiciel FrontPage\texttrademark{} 2002, une clause de censure
interdisant \og \textit{d'utiliser {[}certains composants de ce
logiciel{]} sur un site qui dénigre Microsoft, MSN, MSNBC, Expedia
ou leurs produits ou services {[}...{]} ou incite au racisme, à la
haine ou à la pornographie} \fg{}.%
}. La très grande liberté de rédaction de ces licences a permis l'expérimentation
de multiples modèles économiques, comme l'illustrent deux sous-classes
des licences privatives~: les licences de type \og partagiciel \fg{},
ou \og \foreignlanguage{american}{\textit{shareware}} \fg{}, qui
autorisent la libre redistribution ---\,selon le modèle dit
du \og marketing viral \fg{}\,--- mais conditionnent l'usage
au paiement d'une redevance, et les licences de type \og gratuiciel \fg{},
ou \og \foreignlanguage{american}{\textit{freeware}} \fg{}, qui
autorisent l'usage gratuit mais interdisent souvent la redistribution
---\,l'obligation de téléchargement auprès du site de l'éditeur
permettant à celui-ci d'accumuler des informations commerciales. Citons
également pour l'anecdote les \og carticiels \fg{}, qui demandent
que l'utilisateur envoie une carte postale de remerciement à l'auteur,
ou encore les \og cariticiels \fg{}, qui requièrent l'envoi d'une
contribution à une organisation caritative laissée au libre choix
de l'utilisateur.

\medskip{}

Cependant, bien avant l'apparition des éditeurs, d'autres modes de
production mettaient en œuvre une mutualisation du coût de développement
entre les usagers~: il s'agissait des clubs d'utilisateurs, encouragés
et parfois même financés par les constructeurs informatiques. Leurs
membres étaient très enclins à partager les outils logiciels qu'ils
développaient, car cela leur permettait de mutualiser leur charge
de travail par rapport à leur objectif~: faire fonctionner de façon
optimale leur système informatique.

L'apparition du modèle d'éditeur commercial sembla sonner le glas
de ces clubs, perçus comme des concurrents, voire des parasites, sur
ce marché naissant. L'une des illustrations les plus frappantes de
cette divergence de pensée est la lettre ouverte écrite par Bill Gates
aux \og hobbyistes \fg{}, en date du 3~février 1976%
\footnote{Cf.~: \href{http://en.wikipedia.org/wiki/File:Bill_Gates_Letter_to_Hobbyists.jpg}{http://en.wikipedia.org/wiki/File:Bill\_{}Gates\_{}Letter\_{}to\_{}Hobbyists.jpg}~.%
}, dans laquelle le jeune entrepreneur se désole du faible taux d'achat
de son interpréteur Basic auprès des possesseurs d'ordinateurs cibles.

Cependant, suivant un phénomène de co-évolution classique, tandis
que se développaient les éditeurs privatifs, un autre système de protection
se mit parallèlement en place dans le milieu des \og hobbyistes \fg{}.
Il prit la forme de \foreignlanguage{american}{\textit{licenses}}
spécifiques, dites \og libres \fg{}, s'appuyant sur le droit du
copyright récemment acquis par les logiciels. La construction de ce
corpus juridique et théorique fut elle aussi progressive. Nous en
retiendrons quelques dates clés, telles que la création de la \foreignlanguage{american}{Free
Software Foundation} en octobre~1985, et la publication de la version~1
de la \foreignlanguage{american}{\textit{General Public License}}
en février~1989. Tout comme les licences privatives, les licences
libres furent favorablement reçues par les juges nationaux%
\footnote{Cf.~: CA Paris, pôle 5, ch. 10, 16~sept. 2009, SA EDU 4
  c/~Association AFPA~; TGI Chambéry, 15~nov. 2007, Espaces et Réseaux
  Numérique c/~Conseil général de Savoie et Université de Savoie~; TGI
  Paris, 28 mars 2007, Educaffix c/~Cnrs, Université Joseph Fourier
  \textit{et al}.%
}.

Les deux écosystèmes coexistent toujours, chacun occupant les environnements
au sein desquels il est le plus efficace, la frontière entre les deux
univers pouvant également évoluer au gré des législations favorables
ou contraires à l'un ou l'autre modèle.

\section{Le numérique irrigue la société}

On peut isoler, dans le continuum historique de l'informatique, des
dates charnières associées à l'apparition de produits représentatifs
de changements de paradigmes. On pense par exemple à l'adoption du
Minitel en France, des interfaces graphiques, etc. Il en est de même
au niveau du droit~: l'évolution, nécessairement lente, des mentalités
et des pratiques se cristallise en un certain nombre de textes faisant
référence par la suite. Nous allons, dans cette section, essayer d'en
évoquer quelques uns.

\subsection{Le statut des données personnelles}

C'est à la suite d'un projet de fusion de différents fichiers contenant
des données personnelles concernant les citoyens français, que l'informatique
fit pour la première fois intrusion dans le champ juridique. L'Insee
et plusieurs services administratifs ayant eu le projet d'interconnecter
leurs fichiers de données personnelles au sein du système Safari (pour
\og Système automatisé pour les Fichiers Administratifs et le Répertoire
des Individus \fg{}), de nombreuses voix s'élevèrent contre cette
initiative jugée liberticide. La plus emblématique fut sans doute
celle de Jacques Fauvet, dans sa célèbre chronique du Monde du 21~mars
1974~: \og \textit{Safari ou la chasse aux français} \fg{}.

Le gouvernement, ayant pris conscience du danger potentiel de ce projet,
le fit enterrer et réagit en promulguant deux ans plus tard la loi
\og Informatique et libertés \fg{} du 6~janvier 1978, l'une des
premières législations nationales portant sur les données personnelles
a entrer en vigueur au monde%
\footnote{Le premier État à légiférer sur cette question fut la Suède (1973),
après le land de Hesse (Allemagne, 1971).%
}.

Ce texte a pour but d'éviter ou de contrôler la constitution de fichiers
directement ou indirectement nominatifs qui soient dangereux pour
les personnes concernées du fait des informations recueillies et conservées,
ou qui pourraient le devenir s'ils étaient détournés de leur finalité
première. Sont naturellement soit interdits, soit soumis à déclaration
préalable, la constitution de fichiers portant sur la vie privée,
les opinions politiques, philosophiques ou religieuses, les mœurs
ou encore l'état de santé.

La surveillance de la bonne application de ce texte fut confiée à
une Autorité administrative indépendante, la CNIL, qui fut la première
du genre en Europe%
\footnote{Il en existe aujourd'hui une trentaine. Au sein de l'Union européenne,
elles collaborent au sein du groupe de travail \og article 29 \fg{}
sur la protection des données, institué par l'article~29 (d'où le
nom) de la directive 95/46/CE du Parlement européen et du Conseil
du 24~octobre 1995, relative à la protection des personnes physiques
à l'égard du traitement des données à caractère personnel. Cette directive
reprit au niveau communautaire nombre des dispositions prises par
les États précurseurs.%
}. Cette autorité s'est vue confier un pouvoir d'avis simple ou conforme%
\footnote{Un avis est dit conforme si la puissance publique ne peut y déroger.%
} selon le traitement envisagé. La CNIL n'a cependant jamais utilisé
ce pouvoir d'opposition, préférant adjoindre des réserves quant aux
traitements de données envisagés.

La réforme de~2004 est venue modifier sensiblement l'équilibre législatif
antérieur, en supprimant le pouvoir d'opposition de la CNIL, qu'elle
aurait peut être enfin utilisé pour s'opposer à certains traitements
qui ont été mis en œuvre par la suite. Ce texte avait l'ambition affichée
de \og toiletter \fg{} la loi pour lui permettre de répondre au
défi de l'utilisation massive du réseau et de l'inévitable internationalisation
des échanges. Force est pourtant de reconnaître qu'une réponse législative
nationale n'a qu'une portée très limitée sur un réseau mondial par
définition.

\subsection{Le statut des systèmes informatiques}

Au delà de la question des données personnelles s'est également posée
celle de la protection des données en général, ainsi que des systèmes
informatiques qui les hébergent. Les conséquences d'actes malveillants
tels que l'entrave au fonctionnement de systèmes informatiques ou
la falsification de leurs données sont d'une gravité potentielle telle
qu'ils devaient assurément relever du droit pénal. Cependant, le législateur
se trouvait fort démuni, car en matière de droit pénal prévaut le
principe de l'interprétation stricte~: le juge ne peut étendre à
sa guise la portée d'articles existants pour réprimer des crimes et
délits voisins. Dès lors, ni les dispositions contre la violation
de domicile, ni celles réprimant les faux en écriture publique ou
privée, ne pouvaient être invoquées.

C'est pour remédier à ce vide juridique que fut votée la Loi du 5~janvier
1988, dite \og loi Godfrain \fg{}. Cette loi créa une notion juridique
nouvelle, le \og \textit{système de traitement automatisé de données} \fg{},
ainsi que plusieurs délits spécifiques~: l'accès frauduleux à un
tel système, l'atteinte au fonctionnement de ce système, et les atteintes
(ajouts, modifications, suppressions) aux données hébergées par ce
système. Les tentatives de ces actes tombent également sous le coup
de la loi.

Afin que cette loi ne soit pas rendue obsolète par l'évolution de
la technique, le législateur se garda bien de fournir une quelconque
définition de la notion de \og \textit{système de traitement automatisé
de données} \fg{}. Tout dispositif cabable d'héberger des données
est potentiellement protégé, pour peu que les élements constitutifs
de l'infraction soient réunis.

La spécificité de ces délits n'échappa pas au législateur, qui regroupa
les articles de la loi \og Godfrain \fg{} au sein d'un nouveau chapitre~III
du Code pénal. Ce chapitre, initialement intitulé \og \textit{De
certaines infractions informatiques} \fg{}, fut renommé en \og \textit{Des
atteintes aux systèmes de traitement automatisé de données} \fg{}
dans le nouveau Code pénal de~1992. Le droit de l'informatique s'était
dans l'intervalle bien étoffé, et la précision devenait nécessaire.

\subsection{L'interopérabilité et les formats}

Une première caractéristique du marché du logiciel est sa très grande
volatilité intrinsèque. Bien que le logiciel soit un produit immatériel,
et donc en théorie inusable, la durée de vie commerciale moyenne d'un
logiciel ne dépasse pas quelques années. Les causes en sont multiples,
et tiennent tant à l'obsolescence rapide des matériels sous-jacents,
supplantés chaque année par des versions améliorées et potentiellement
incompatibles, qu'à l'évolution rapide des diverses couches logicielles.
Ce phénomène est prééminent dans le monde du jeu vidéo, pour lequel
chaque œuvre logicielle est indépendante des autres et poursuit une
carrière commerciale propre.

Une deuxième caractéristique, qui s'oppose à la première, est la très
grande prééminence des effets de réseau%
\footnote{On appelle ainsi le fait que la valeur d'un bien augmente avec le
nombre de personnes qui l'utilisent.%
}. L'une des principales sources d'effet de réseau dans le monde numérique
est l'encodage des informations au sein de protocoles d'échange et
de formats de fichiers spécifiques à un logiciel donné. Maintenir
secrets ces protocoles et formats favorise la création de marchés
captifs. L'exemple classique est celui des outils de bureautique~:
plus on utilise un logiciel donné pour créer des documents que seul
ce logiciel sait lire, et plus on incite les personnes avec qui l'on
est en relation, ainsi que soi-même, à utiliser ce logiciel afin de
pouvoir accéder aux documents déjà créés. On observe également cet
effet au niveau des outils de communication%
\footnote{Comme par exemple le logiciel Skype\texttrademark{}, dont le protocole
de communication n'est pas documenté.%
} et des sites centralisés hébergeant des réseaux sociaux.

L'informatisation croissante de la société passe par la possibilité
d'échanger, entre logiciels, des données de sources différentes. Il
est également nécessaire de pouvoir transférer ses données d'un logiciel
à son successeur. Ces conditions sont indispensables à l'existence
d'une concurrence libre et non faussée dans le secteur du logiciel,
qui faute de cela serait le havre des marchés captifs et des ventes
liées.

Les auteurs de logiciels n'ayant pas le temps de réaliser tous les
modules de conversion entre formats différents, ou ne souhaitant simplement
pas le faire à un coût raisonnable afin de garder captive leur clientèle,
le législateur européen a introduit une disposition originale au sein
de la directive 91/250/CE déjà mentionnée~: le droit de rechercher
l'interopérabilité.

Le droit d'auteur adapté interdit à quiconque d'effectuer la rétro-ingénierie
par décompilation d'un logiciel existant. Les compétiteurs d'un logiciel
initial peuvent certes observer le fonctionnement externe de ce logiciel
(art. L.122-6-1 III\textdegree{} du CPI). En revanche, ils ne peuvent
pas reprendre à leur compte l'expertise de codage acquise par les
auteurs initiaux, et doivent supporter eux aussi les coûts et les
temps de spécification et de codage de leur propre logiciel. Faute
de cela, on autoriserait la recopie servile par décompilation de la
forme de l'œuvre logicielle, action assimilable au plagiat littéraire.
L'interdiction de la décompilation maintient tous les compétiteurs
sur un pied d'égalité, en garantissant l'avantage compétitif du premier
entrant.

La seule exception à cette interdiction concerne les opérations visant
à \og \textit{obtenir les informations nécessaires à l'interopérabilité
d'un logiciel créé de façon indépendante avec d'autres logiciels} \fg{}
(art. L.122-6-1 IV\textdegree{}). Cette disposition, très encadrée
afin de ne pas être dévoyée, est une spécificité européenne. Elle
permet aux usagers d'un logiciel de pouvoir toujours accéder légalement
à leurs données, même en cas de disparition ou du refus de l'éditeur
de ce dernier.

La question qui se pose alors est celle de la pertinence économique
des formats de données fermés, en regard de l'intérêt général. Dès
le moment où la rétro-ingénierie à fin d'interopérabilité d'un format
de fichier fermé a été effectuée par l'auteur d'un logiciel libre,
les spécifications de ce format deviennent accessibles à tous~; le
format devient, en quelque sorte, un format ouvert%
\footnote{Ceci n'est pas tout à fait exact, en ce qui concerne la question centrale
de la gouvernance vis-à-vis de l'évolution des caractéristiques de
ce format.%
}. On pourrait à bon droit considérer que la loi devrait explicitement
obliger les auteurs de logiciels à offrir à leurs usagers la description
du format qui contient les données qui leur appartiennent ou, alternativement,
à garantir l'exportation de leurs données sous un tel format documenté.

Le travail du législateur a, en la matière, un goût d'inachevé. Si
la notion de standard ouvert a été définie par la Loi du 21~juin
2004, qui dispose en son Article~4 que~: \og \textit{On entend
par standard ouvert tout protocole de communication, d'interconnexion
ou d'échange et tout format de données interopérable et dont les spécifications
techniques sont publiques et sans restriction d'accès ni de mise en
œuvre} \fg{}, l'interopérabilité ne l'a pas été, laissant la porte
ouverte aux divergences d'interprétation. La promotion de l'un et/ou
de l'autre par l'action publique n'a pas non plus fait l'objet d'une
législation, comme c'est pourtant le cas dans d'autres pays.

\subsection{L'émergence des réseaux\label{sub:L'=0000E9mergence-des-r=0000E9seaux}}

Au tournant des années~1980, les moyens d'accès à distance se démocratisèrent.
La majorité des particuliers utilisaient des modems pour se connecter
aux systèmes distants à travers le réseau téléphonique, avant qu'Internet
ne perce auprès du grand public.

L'informatisation croissante de la société s'accompagna d'un double
changement de paradigme quant à la place de l'ordinateur dans la société.
Le premier est sa transformation d'un outil d'information en un outil
de communication. L'ordinateur n'est plus un simple serveur d'informations
provenant d'une source autorisée, mais permet aussi l'échange de contenus
entre usagers. Cet aspect de la \og \textit{révolution télématique}%
\footnote{\textit{France Télécom~: un opérateur de réseau devient un acteur
de la communication}, Jean-Marie Charon, Réseaux n\textdegree{}~37,
CNET et TIS, vol~2(1), 1989.\hfill~\linebreak Cf.~: \href{http://revues.mshparisnord.org/lodel/disparues/docannexe/file/103/tis_vol2_n1_2_29_50.pdf}{http://revues.mshparisnord.org/lodel/disparues/docannexe/file/103/tis\_{}vol2\_{}n1\_{}2\_{}29\_{}50.pdf}~.%
} \fg{} a été popularisé, en France, par les messageries. Le deuxième
changement est l'interconnexion des équipements au sein de réseaux
de plus en plus vastes, jusqu'au \og réseau des réseaux \fg{} que
constitue Internet. On n'accède plus à un ordinateur, on accède à
un réseau, ce dernier donnant accès à des ressources de moins en moins
individualisées. La vision de John Gage, \og \foreignlanguage{american}{\textit{the
network is the computer}} \fg{}, prend forme avec l'émergence de
l'\og informatique en nuage \fg{}.

L'externalisation des infrastructures matérielles et logicielles a
conduit à une profonde modification des licences logicielles. Du fait
de l'utilisation des logiciels \og en tant que service \fg{} (\og SaaS \fg{},
pour \og \foreignlanguage{american}{\textit{Software as a Service}} \fg{}),
les droits concédés ne concernent plus que la simple utilisation,
tous les autres droits (correction des bogues, copie de sauvegarde,
etc.) devenant caducs du fait que le logiciel proprement dit n'est
plus accessible aux usagers.

Ce phénomène a également impacté les licences libres. Le mécanisme
d'adhésion aux termes de ces licences, qui était déclenché lors du
téléchargement ou bien de la redistribution d'un logiciel, devenait
inopérant lorsque le logiciel était simplement utilisé à distance.
Au contraire même, il était clairement stipulé que l'usage du logiciel
appartenait à la sphère privée. Ceci permit à de nombreux prestataires
d'offrir des services basés sur des logiciels libres sans reverser
à la communauté les modifications auxquelles les utilisateurs auraient
pu avoir accès si le logiciel avait été installé sur leur ordinateur%
\footnote{Cet échappatoire aux termes de la licence est appelé \og \foreignlanguage{american}{\textit{ASP
loophole}} \fg{}.%
}.

C'est pour prendre en compte ce phénomène que la société Affero diffusa
en mars~2002 la première licence libre déclenchée par l'utilisation
à distance, l'AGPL~v1%
\footnote{\selectlanguage{american}%
\textit{Affero General Public License, version~1}\foreignlanguage{french}{.
Affero.inc. Cf.~: \href{http://www.affero.org/oagpl.html}{http://www.affero.org/oagpl.html}~.}\selectlanguage{french}%
}, élaborée avec l'aide de la \foreignlanguage{american}{Free Software
Foundation}.

Cependant, ces licences ne règlent absolument pas le statut des données.
Avec une licence déclenchée par l'usage, un utilisateur peut effectivement
consulter le code source du logiciel qu'il utilise à distance, mais
il ne peut aucunement modifier l'exemplaire du prestataire pour l'adapter
à ses besoins. On a même pu voir des prestataires \og mutiler \fg{}
des logiciels libres afin d'empêcher les usagers d'exporter les données
qu'ils avaient introduites dans le système distant, alors que ces
fonctionnalités étaient présentes dans la version d'origine du logiciel%
\footnote{Ce fut le cas par exemple pour les services offerts par la société
Xooit. Cf.~: \href{http://www.xooit.com/}{http://www.xooit.com/}~.%
}.

Alors que de plus en plus de contenus accessibles à travers Internet
sont auto-produits, la liberté d'accès à ses propres données reste
à construire%
\footnote{On suivra avec intérêt les développements du procès que Max Schrems,
entre autres, a intenté à FaceBook\texttrademark{} quant à l'accès
à ses données personnelles.%
}. On peut la considérer tant comme une question d'interopérabilité,
du point de vue des traitements, que d'accès aux données dont on est
le créateur voire l'auteur si elles sont de nature littéraire et/ou
artistique.

\subsection{Le droit d'Internet}

Internet est unique par définition, puisque c'est le \og réseau des
réseaux \fg{}. L'émergence d'Internet comme réseau de communication
universel, qui absorbe maintenant le réseau télévisuel après avoir
absorbé le réseau téléphonique, conduit la puissance publique à s'interroger
régulièrement sur le statut de ce réseau. La nouveauté effraie, y
compris le législateur, ce qui fait que les premières tentatives législatives
ont toutes tenté de réduire les libertés sur le réseau, plutôt que
de garantir les droits des usagers et prestataires. L'incompréhension
du fonctionnement du réseau a également conduit à une confusion dommageable
entre les rôles des fournisseurs d'accès, des hébergeurs, des éditeurs,
des auteurs et des usagers, qui perdure encore partiellement aujourd'hui.

L'origine de cette incompréhension repose sur la volonté inflexible
du juge de trouver un responsable aux agissements contraires au droit
commis sur les réseaux, quel qu'il soit. C'est ainsi qu'en~1996,
alors que des images à caractère pédophile avaient été diffusées sur
des serveurs de \og news \fg{}, ce furent deux fournisseurs d'accès
à Internet (FAI), Worldnet et Francenet, qui furent mis en accusation.
En dépit de l'évidente erreur de cible, il fallut plusieurs années
avant que les poursuites ne soient abandonnées.

La deuxième victime expiatoire, faute de retrouver la personne à l'origine
d'une mise en ligne délictueuse, fut l'hébergeur. C'est ainsi que,
dans l'affaire Estelle Haliday c/~Valentin Lacambre, c'est ce dernier,
responsable de l'hébergeur Altern, qui fut condamné, tant en première
instance qu'en appel. D'autres gérants de forums furent mis en garde
à vue dans des affaires similaires. Face à la faiblesse juridique
de ces décisions, plusieurs lois furent proposées afin d'encadrer
la responsabilité juridique des intermédiaires techniques. Ce fut
le cas de la Loi Fillon de 1997, portant obligation aux hébergeurs
de surveiller leurs hébergés, ou de la loi Jospin de 2000, aux dispositions
similaires. Toutes deux furent, fort heureusement, invalidées par
le Conseil constitutionnel. Il fallut du temps au législateur pour
rédiger une loi qui passe sans trop de dégâts sous les fourches caudines
du Conseil. Tel fut le cas de la Loi du 21~juin 2004, qui définit
une responsabilité atténué des hébergeurs~: ces derniers ne sont
incriminés que s'ils ne font pas diligence à supprimer un contenu
jugé illicite, sur injonction judiciaire. De même, les FAI doivent-ils
disposer de moyens de filtrage mobilisables en cas de décision de
justice. Ce fut le cas dans l'affaire \og Yahoo US \fg{}, en 2000,
dans laquelle il fut enjoint aux principaux FAI français de filtrer
le site permettant d'acheter des objets promouvant l'idéologie nazie,
sur la base du trouble manifeste à l'ordre public.

\medskip{}

Toutes les interventions publiques n'ont cependant pas visé à restreindre
les libertés des usagers. Le Programme d'action du gouvernement pour
la société de l'information (PAGSI) de 1997, adopté en 1998 par le
Comité interministériel pour la société de l\textquoteright{}information
créé pour sa mise en œuvre, affiche des objectifs ambitieux pour l'époque.
Il introduit la notion d'accès universel à Internet (tous les Français
devaient pouvoir y avoir accès à des conditions tarifaires équivalentes
quel que soit leur lieu de résidence), et stimule la création de points
d'accès publics pour les personnes ne disposant pas d'un ordinateur
chez elles. Il encourage également la mise en ligne des données publiques%
\footnote{En rupture avec la pratique en vigueur, qui avait été de concéder
un monopole exclusif de mise en ligne sur serveur Minitel payant des
textes de loi, marché lucratif concédé à la société OR Télématique
(la bien nommée).%
}.

\medskip{}

En matière d'Internet, les questions juridiques sont légions. On peut
citer, par exemple, celles de la neutralité des réseaux, du droit
à l'oubli, etc. Cependant, on sort ici quelque peu du périmètre historique
pour rentrer dans l'actualité brûlante. Une évocation de ces questions
peut néanmoins sembler pertinente dans une démarche explicative vis-à-vis
du grand public.

\section{Droit et muséographie du logiciel}

\subsection{Droit du logiciel et muséographie}

Le rattachement du logiciel au droit d'auteur, s'il l'a fait entrer
dans un périmètre juridique bien connu des conservateurs, a eu pour
conséquence funeste une extension considérable de la durée des droits
qui le couvrent~: d'une durée fixe de 25~ans, on est passé à 70~ans
après l'année de la première parution pour les œuvres collectives,
et à 70~ans après le décès du dernier auteur survivant pour les œuvres
de collaboration (art. L.123-2 et~-3 du CPI).

Ces durées sont sans commune mesure avec la rapidité de l'évolution
dans ce secteur. Celle-ci rend très vite obsolètes les logiciels,
ce qui les fait rapidement entrer dans l'histoire. Elle induit également
un renouvellement perpétuel du tissu industriel, les entreprises titulaires
des droits sur ces logiciels pouvant disparaître ou se restructurer
en l'espace de quelques années.

La presque totalité des œuvres logicielles susceptibles de faire l'objet
d'une conservation à titre historique appartient donc à la catégorie
des œuvres dites \og indisporphelines \fg{}. Ce mot-valise englobe
les œuvres \og indisponibles \fg{}, c'est-à-dire qui ne sont plus
diffusées par un éditeur encore existant, et \og orphelines \fg{},
c'est-à-dire dont le titulaire des droits n'est plus connu, comme
par exemple dans les cas où l'entreprise cessionnaire des droits a
disparu ou lorsque l'auteur est décédé sans ayant-droit connu).

\medskip{}

La Cour de justice de l'Union européenne a récemment apporté une clarification
bienvenue sur la transmission des licences logicielles, en son arrêt
du 3~juillet dernier%
\footnote{Arrêt C-128/11, UsedSoft GmbH~/ Oracle International Corp.\hfill~\linebreak
Cf.~: \href{http://curia.europa.eu/jcms/upload/docs/application/pdf/2012-07/cp120094fr.pdf}{http://curia.europa.eu/jcms/upload/docs/application/pdf/2012-07/cp120094fr.pdf}~.%
}. Ce dernier stipule qu'un éditeur de logiciel ne peut s'opposer à
la revente (ni donc à la cession) de licences \og d'occasion \fg{}
de son logiciel, que celles-ci soient attachées à un support physique
ou bien aient été acquises au travers d'Internet.

La personne ayant obtenu un logiciel de seconde main est donc pleinement
en droit d'effectuer toutes les actions que le CPI autorise, incluant
la réalisation d'une copie de sauvegarde à partir d'un support dégradé
(tels que les anciens supports magnétiques). Le plein exercice de
ce droit nécessite cependant de pouvoir prouver que le logiciel a
bien été acquis de façon licite, c'est-à-dire que l'on puisse reconstituer
la chaîne des cessions qui l'auront amené entre les mains du conservateur,
ce qui peut constituer en pratique un obstacle souvent insurmontable.

En revanche, cette décision ne s'applique pas aux logiciels dont la
licence est limitée dans le temps (location), et pour laquelle il
sera nécessaire d'obtenir une nouvelle autorisation du titulaire des
droits%
\footnote{Ceci étant évidemment problématique pour les œuvres orphelines.%
}. Les conservateurs du patrimoine logiciel seront confrontés de façon
croissante à ce problème, ce mode de diffusion étant de plus en plus
utilisé par les éditeurs. Le maintien d'un serveur de licences fonctionnel
sera qui plus est nécessaire à la mise en œuvre des logiciels qui
y ont recours.

Un amendement à la loi \og Création et Internet \fg{} (dite \og Hadopi \fg{})
de~2009 a précisé les conditions dans lesquelles les musées et bibliothèques
peuvent effectuer leurs actions de conservation et de présentation
des œuvres au public%
\footnote{Cet amendement autorise~: \og {[}...{]} \textit{la reproduction
et la représentation d\textquoteright{}une œuvre faisant partie de
leur collection effectuée à des fins de conservation ou destinée à
préserver les conditions de sa consultation sur place à des fins de
recherche ou d\textquoteright{}études privées dans les locaux de l\textquoteright{}établissement
et sur des terminaux dédiés par des bibliothèques accessibles au public,
par des musées ou par des services d\textquoteright{}archives, sous
réserve que ceux-ci ne recherchent aucun avantage économique ou commercial} \fg{}.
Cf.~: \href{http://scinfolex.wordpress.com/2009/05/17/une-nouvelle-formulation-pour-lexception-bibliotheques-dans-la-loi-hadopi/}{http://scinfolex.wordpress.com/2009/05/17/une-nouvelle-formulation-pour-lexception-bibliotheques-dans-la-loi-hadopi/}~.%
}. Cet amendement, qui autorise \og \textit{la représentation} {[}...{]}
\textit{sur des terminaux dédiés} \fg{}, peut être interprété comme
autorisant la présentation en fonctionnement de logiciels sur des
ordinateurs spécialement disposés à cette fin.

\subsection{Muséographie du droit du logiciel}

Les évolutions juridiques sont le reflet des évolutions sociétales
induites par la \og révolution numérique \fg{}. Il s'agit pour le
législateur d'accompagner l'évolution des pratiques en offrant un
cadre régulateur apte à favoriser les pratiques considérées comme
vertueuses et à décourager les pratiques considérées comme nuisibles.

La présentation du cheminement du droit du numérique doit donc selon
nous accompagner la présentation des solutions techniques (tant matérielles
que logicielles) et des usages, afin de refléter le plus complètement
possible l'esprit de chacune des époques traversées.

La nécessité de l'évolution du droit apparaîtra clairement par la
mise en évidence du changement de paradigme économique induit par
la numérisation de l'information. La possibilité d'abstraire l'information
de tout support%
\footnote{Bien que nombre de pièces de musée soient en fait des dispositifs
matériels.%
} induit des conditions de production et d'échange des savoirs radicalement
nouvelles. La copie s'effectuant à coût nul (l'action de copie ne
modifie que marginalement la consommation en ressources du matériel
sous-jacent), les accès aux réseaux étant eux aussi forfaitisés, plus
aucune barrière ne vient empêcher la micro-création de valeur ajoutée,
rendant possible le travail collaboratif à grande échelle. L'abondance
d'informations qui en résulte, et la possibilité pour quiconque d'en
être producteur, ne peuvent qu'entrer en conflit avec la vision ancienne
des droits d'auteur, pensée pour un monde où la rareté était la norme.

La présentation de l'avancée du droit et des usages pourra se faire
tant par des frises et panneaux synthétiques que par l'installation
d'encarts dédiés accompagnant les diverses pièces (matérielles et
logicielles).

La documentation contractuelle afférente aux matériels et logiciels
est encore moins accessible que ces derniers, car souvent gérée par
des services différents. Son cycle de vie n'est pas non plus le même~:
elle est en général conservée pour archive alors que le matériel est
déclassé et récupéré par des tiers, puis détruite à la fin de la durée
légale de conservation. Des actions de collecte spécifique devront
être entreprises si nécessaire.

\section*{Conclusion}

La muséographie de l'informatique est confrontée à un double défi
en ce qui concerne le droit du numérique. Le premier est la présentation
de ce secteur, nécessaire à la compréhension du statut unique du logiciel
dans l'histoire humaine et de l'impact de la révolution numérique
sur l'ensemble de la société. Le deuxième découle des contraintes
que ce droit fait lui-même peser sur la préservation et la présentation
des œuvres, et en particulier des œuvres logicielles. Autant la possession
d'un exemplaire de machine ou d'ouvrage vaut droit d'usage et de présentation%
\footnote{Mais non de \og représentation \fg{} au sens où l'entend le droit
d'auteur dans le cas d'une œuvre telle qu'une pièce de théâtre. Ces
restrictions interdisent également la diffusion publique d'un vidéogramme.%
}, autant l'acquisition du support d'un exemplaire de logiciel ne garantit
par elle-même aucun droit sur ce dernier. Un travail de \og restauration
juridique \fg{} doit donc être entrepris afin de garantir que chaque
pièce puisse être présentée au public sans risque pour l'exposant.
\end{document}